\documentclass{article}

\usepackage{natbib}
\setcitestyle{numbers}
\setcitestyle{square}

\usepackage{amsmath,graphicx}
\usepackage{subcaption}
\usepackage[T1]{fontenc}
\usepackage{amssymb}
\usepackage{enumitem}
\usepackage{multirow}
\usepackage[table]{xcolor} 
\usepackage{url}
\usepackage{makecell}
\usepackage{comment}
\usepackage{highlight}
\usepackage{fancybox}
\usepackage{tcolorbox}
\usepackage{float}
\usepackage{stfloats}
\usepackage[font=footnotesize]{caption}
\newtcolorbox{mytcolorbox}{fontupper=\footnotesize,boxrule=1pt,boxsep=0pt,left=4pt,right=4pt,top=4pt,bottom=4pt}
\renewcommand{\opacity}{50}
\usepackage[belowskip=-10pt,aboveskip=2pt]{caption}
\newcolumntype{Y}{>{\centering\arraybackslash}X}
\renewcommand{\arraystretch}{1.5}

\newlength\myindent
\def\Width{0\kern2\tabcolsep\ldots\kern1\tabcolsep0}
\usepackage{etoolbox}
\newcommand{\zerodisplayskips}{%
  \setlength{\abovedisplayskip}{3pt}
  \setlength{\belowdisplayskip}{3pt}
  \setlength{\abovedisplayshortskip}{3pt}
  \setlength{\belowdisplayshortskip}{3pt}}
\appto{\normalsize}{\zerodisplayskips}
\appto{\small}{\zerodisplayskips}
\appto{\footnotesize}{\zerodisplayskips}

\usepackage[final]{neurips_2024}

\usepackage[utf8]{inputenc} % allow utf-8 input
\usepackage[T1]{fontenc}    % use 8-bit T1 fonts
\usepackage{hyperref}       % hyperlinks
\usepackage{url}            % simple URL typesetting
\usepackage{booktabs}       % professional-quality tables
\usepackage{amsfonts}       % blackboard math symbols
\usepackage{nicefrac}       % compact symbols for 1/2, etc.
\usepackage{microtype}      % microtypography
\usepackage{xcolor}         % colors

\title{Device-Directed Speech Detection for Follow-up Conversations Using Large Language Models}

\author{Ognjen (Oggi) Rudovic$^*$, Pranay Dighe$^*$, Yi Su$^*$, Vineet Garg, Sameer Dharur, \\ {\bf Xiaochuan Niu, Ahmed H. Abdelaziz,  Saurabh Adya, Ahmed Tewfik}\\{\bf Apple}}
\begin{document}
\maketitle
\def\thefootnote{*}\footnotetext{Equal Contribution. Correspondence: {\it \{orudovic,yi\_su\}@apple.com}}

\begin{abstract}
Follow-up conversations with virtual assistants (VAs) enable a user to seamlessly interact with a VA without the need to repeatedly invoke it using a keyword (after the first query). Therefore, accurate Device-directed Speech Detection (DDSD) from the follow-up queries is critical for enabling naturalistic user experience. To this end, we explore the notion of Large Language Models (LLMs) and model the first query when making inference about the follow-ups (based on the ASR-decoded text), via prompting of a pretrained LLM, or by adapting a binary classifier on top of the LLM. In doing so, we also exploit the ASR uncertainty when designing the LLM prompts. We show on the real-world dataset of follow-up conversations that this approach yields large gains (20-40\% reduction in false alarms at 10\% fixed false rejects) due to the joint modeling of the previous speech context and ASR uncertainty, compared to when follow-ups are modeled alone.
\end{abstract}

\vspace{-2mm}
\section{Introduction}
\label{sec:intro}
\begin{figure}[th]
    \captionsetup{font=footnotesize}
    \centering
    \includegraphics[width=0.6\linewidth]{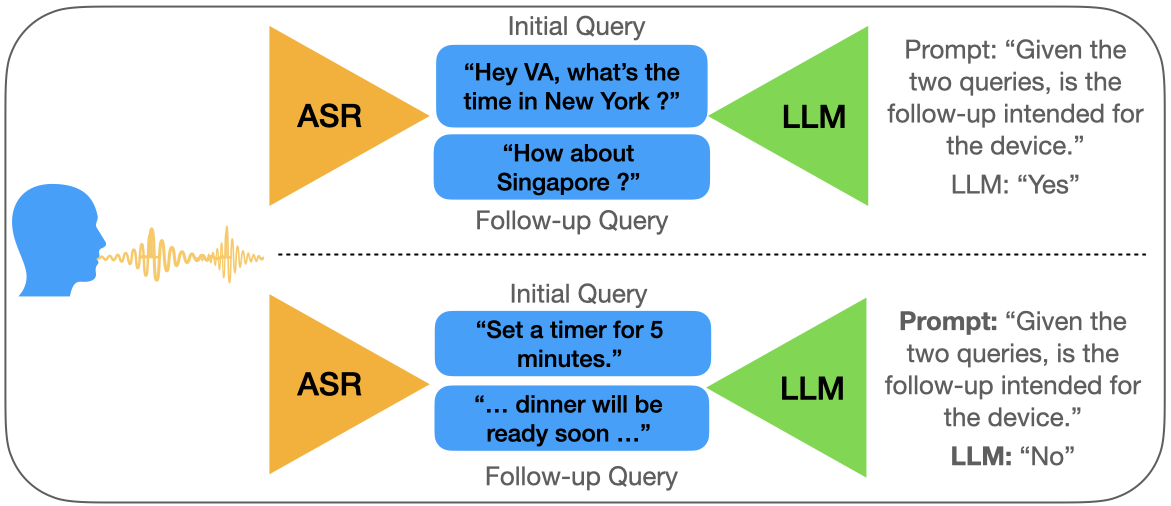}
    \caption{Follow-up conversations: the pair of user queries are first processed by an ASR system, which outputs the text transcriptions of the user's speech. The joint ASR transcription of the initial and follow-up queries are input to the LLM that detects if the latter is directed to the Virtual Assistant (VA).}
    \label{fig:motivation}
\end{figure}

\vspace{-1mm}
%intro about B2B
Virtual assistants (VAs) are at the core of smart devices (e.g., mobile phones, smart speakers, wearables, etc.) as they aim to enable a naturalistic voice-based interaction between a user and a device. For VAs to respond to the user requests reliably, they need to infer whether the user is talking to the device or not. For instance, the user could be talking to someone else, and/or there could be side-speech conversations, background noise, etc. Therefore, classifying accurately if the user's speech is device-directed is critical for providing relevant responses, and to avoid interfering with the user’s interactions which are not device-directed, i.e., intended for the VA. This task is often referred to as the device-directed speech detection (DDSD)~\cite{dighe2024llm,streaming-on-device,Mallidi2018, Gillespie2020}. Most existing works on DDSD focus on detection from single queries of the user, often beginning with a wakeword (e.g., ``Hey Google'', ``Hey Siri'', ``Alexa'', and so on). Such isolated utterances are usually a \textit{complete} question or a task request from a user to the VA, and often do not require additional context to determine if the speech is the VA directed. In this work, we address the task of DDSD in follow-up conversations, where the user's first query starts with a wakeword (that is easier to detect with high accuracy by existing systems for the wakeword detection~\cite{sainath2015convolutional}), potentially followed by another query (termed as the ``follow-up''), as a continuation of the conversation with the VA. The follow-ups by design do not require the wakeword, and, therefore, classifying them correctly is far more challenging (see Fig.~\ref{fig:motivation}).

Previous approaches to DDSD process isolated utterances (i.e., no previous context is considered) either directly from audio~\cite{streaming-on-device}, text~\cite{data-resource-efficient}, or from intermediate Automatic Speech Recognition (ASR) lattice-based features~\cite{agrawal2020lrnn}. A few recent approaches attempted classification of device-directed speech in the context of natural turn-taking~\cite{Mallidi2018, Gillespie2020,wang2024turn} by exploring various acoustic and lexical features; however, these works do not account for joint modeling of the ASR uncertainty and multi-turn user queries. The approach proposed here is built upon the recent work in~\cite{dighe2024llm} that focuses on the DDSD task from a single query using a combination of off-the-shelf ASR model ($n$-best ASR hypotheses) and LLM. We focus on a more challenging task -- DDSD for follow-ups, where, given the previous query of the user, the goal is to determine if the follow-up query is directed to the device. For this, we propose two LLM-based approaches: (i) prompting-based, where we experiment with direct {\it text-prompting} of a pretrained, and a finetuned LLM (Sec.~\ref{prompt_base}). (ii) classification-based approach, where, instead of parsing the LLM decisions from output text, as in (i), we introduce a classification head on top of the LLM, so the approach can make probabilistic binary decisions for the DDSD (Sec.~\ref{class_base}). Specifically, we simultaneously address two key limitations of the existing systems for DDSD: (i) the lack of context -- where only the single query is modeled; instead, we account for both the initial and follow-up queries, thus, providing valuable information to the LLM when making decisions, and (ii) ASR uncertainty -- while ASR models aim to transcribe speech to text accurately, speech recognition from the real-world speech is often inaccurate and the $1$-best ASR hypothesis acts as an information bottleneck between the ASR and the LLM component. We expand this information bottleneck by exposing the LLM to an $n$-best list of ASR hypotheses, and we do so only for the follow-up utterance to prevent confusing the model, i.e., we focus on the uncertainty in the follow-up only. Apart from direct prompting, we also explore finetuning of Low-Rank Adaptation (LoRA)~\cite{hu2021lora} adapters using training data of the $n$-best list prompts. We use an in-house general-purpose English ASR system and Vicuna~\cite{vicuna2023}, an instruction-tuned LLaMA LLM~\cite{touvron2023llama}. 

\begin{figure}
\begin{mytcolorbox}
\textbf{Prompt:} In this task, we provide a pair of queries made by human in the following format: `Query 1: $<$text$>$ | Query 2: $<$text$>$'. Query 1 is directed toward the voice assistant. Query 2 is the follow-up query made by human. \{\textit{For Query 2, we provided an n-best list of ASR hypotheses for the spoken utterance. Each of the hypothesis is separated by a newline character. The cost of each hypothesis is at the end in the format `[cost]' where a low cost indicates that we are more confident about that ASR hypothesis.}\} Determine whether Query 2 is directed towards a voice assistant or a human being. Typical spoken utterances directed towards the voice assistant are commands to fulfill a task or queries to get some information. Answer only from the following categories [`1', `0'] where `1' indicates that the utterance is directed towards the voice assistant and `0' indicates that the utterance is directed towards a human being. In your answer the last line should contain nothing else but the number `0' or `1'.
\end{mytcolorbox}
\caption{Task-prompt used for Device-directed Speech Detection. Note that for the follow-up query, the text in \textit{italics} is additionally added when including the n-best hypothesis by the ASR system. }
\label{fig:prompt}
\end{figure}

Our experiments show noticeable gains in terms of the DDSD accuracy over the simple approach where only the follow-up query is used as input to LLM. While this is intuitive, we empirically show on the challenging real-world dataset of follow-up conversations that jointly modeling the context (the previous and follow-up query), together with ASR uncertainty, helps to improve the LLM detection accuracy in the range of $\sim$20-40\%, when the classifier is trained on top of the LLM. It also largely outperforms the traditional prompting-based approach without the LLM tuning on the target task. 

%Compared to the approach proposed in ~\cite{dighe2024llm}, which uses the prompting-based approach, we show that the approach proposed here can increase the intent classification accuracy by more than 15\% mainly due to the new classification head in the LLM, as well as the modeling of the context of the previous query (see Sec.~\ref{results} for details).  

%\vspace{-2mm}
\section{Methods}
\label{sec:approach}

The goal is to identify whether a spoken follow-up utterance is directed towards a device (e.g. smartphone) or a human (see Fig.~\ref{fig:motivation}). To prepare the input for the LLM, we take the ASR outputs that come in the form of $1$-best and $n$-best hypotheses, derived from a \textit{lattice} graph comprised of the competing ASR hypotheses under a beam-search decoder~\cite{mohri2002weighted}. The least cost path in the lattice is the $1$-best hypothesis, while the $n$-best list comes from the full lattice (obtained by picking the $n$ least cost paths in the lattice), which better captures the ASR uncertainty. We describe the follow-up query with either 1-best or n-best hypotheses whereas the initial query is only described using the 1-best hypothesis. We concatenate these ASR information from the initial and follow-up query to obtain the \textit{``utterance-prompt''} (see Table \ref{table:nbest_examples}). In the $n$-best lists of the follow-ups, the hypotheses are separated with newline characters and each hypothesis is appended with a \textit{hypothesis-cost} at the end in the format \textit{[cost]}. This cost is the sum of the acoustic- and language-model costs on the arcs along the lattice-path corresponding to the hypothesis~\cite{povey2012lattice}, where a low cost indicates a high posterior probability assigned by the ASR~\cite{dighe2024llm}.  
%\vspace{-2mm}
\begin{table}[h!]
\centering
\renewcommand{\arraystretch}{1}
\resizebox{0.6\columnwidth}{!}{%
\begin{tabular}{lcc}
 \rowcolor{gray!50}
 \hline
utterance-prompt &Initial & Follow-up  \\\hline\hline
1-best & \textit{``Hey VA, play music''} & \textit{``turn it up a bit''}\\\hline
$n$-best & \textit{``Hey VA, play music''} & \makecell{\textit{``turn it up a bit [-81.4]}\\\textit{turn it up a bet [-78.1]}\\\textit{term it up a pit [-75.9]}}\\\hline
Ground-truth & \textit{``Hey VA, play music''} & \textit{``turn it up a bit''}\\\hline
\end{tabular}
}
\caption{Examples of $1$-best vs. $n$-best lists utterance-prompts, with the hypothesis cost shown in the brackets.}
\label{table:nbest_examples}
\end{table}

\subsection{Prompting-based approach} 
\label{prompt_base}
In Fig.~\ref{fig:prompt}, we show an example of the fixed prompt, referred to as the \textit{task-prompt}, that varies based on whether the $1$- or $n$-best hypotheses are used in the follow-up. This task-prompt is further concatenated with the utterance-prompts containing the prompt with a follow-up. We consider two inference strategies: direct prompting of the pretrained LLM (``PromptOnly''), and prompting of a finetuned LLM (``FinetunePrompt'') where the base LLM is finetuned using Parameter-Efficient FineTuning (PEFT) methods~\cite{peft}, specifically LoRA~\cite{hu2021lora} - see Sec.~\ref{llm:finetune}.

\vspace{-2mm}
\subsection{Classification-based approach}
\label{class_base}
Another way to take advantage of the pretrained LLM is to add a classifier head on top of it, and then to finetune the entire model (we refer to this approach as ``ClassHead''). This approach is popularized by BERT~\cite{bert} in prior literature. One of the drawbacks is that fine-tuning the entire LLM along with the classifier head not only requires a large amount of in-domain training data but also makes it difficult to deploy the model in production. To this end, we use the LoRA adapter finetuning again in the context of adding the LLM with a classifier head. This approach is referred to as ``ClassLoRAHead''. The classifier head in both ``ClassHead'' and ``ClassLoRAHead'' approaches is simply a linear layer that maps the last hidden layer output to a one-hot vector representing the device-directed speech label. In our task, the linear layer has 4096-$D$ input, and 2-$D$ output. The standard cross-entropy loss is used to train the LLM (or its LoRA adapters) and the classifier head.

\vspace{-4mm}
\section{Experiments}
\label{sec:experiments}
%\vspace{-2mm}
%In this section, we provide details of models, training, experiments and subsequent analysis.

%\vspace{-2mm}
%\subsection{Models, Finetuning, and Baselines}
\vspace{-2mm}
\subsection{ASR System and Dataset}
%\vspace{-2mm}
The ASR model that generates the $1$- and $n$-best hypothesis for the input queries is based on the E2E-ASR architecture in~\cite{wu2021u2++}, and it comprises of a Conformer~\cite{gulati2020conformer} encoder with a CTC and attention-based decoder. The beam-search decodings from the attention-based decoder are rescored using an external Finite State Transducer-based language model (FST-LM). The ASR lattices obtained from the FST-LM decoding are used to generate the ASR hypotheses. Specifically, we use an internal audio dataset from demographically diverse English native speakers with consent, which is tailored to simulate a naturally occurring multi-turn conversations towards solving a task together (e.g, cross-word puzzles). The participants ask the VA questions, and use the answers to complete the task at hand, while the VA-equipped devices record audio from the participants. The dataset consists of $\sim$19k audio recordings from 1.3k participants. The conversations are clipped into roughly 245k segments, i.e., the pairs of utterances, where the first utterance is always device-directed, and the follow-up can be either directed or undirected. These data are human-annotated in terms of device-directness, and are split into training, validation, and test (70/10/20\%) partitions, with no speaker overlap. The ratio of the device-directed and -undirected examples is $\sim$1:4, with average audio duration of 3.5$\pm$3.25 (mean$\pm$std) seconds.

\begin{table}[!htb]
    \caption{Accuracy of the prompting- and classification-based LLM approaches (in \%). ({\it left}) The upper / lower rows show results for the follow-up query only (no context) / and pairs of queries (context), with the number of the ASR hypotheses ($n$) in the first column. ({\it right}) LLM models with a classification head. The  paired t-test with 95\% confidence intervals reveals statistically significant improvement (*)  when comparing the last two rows (while both approaches bring statistically significant improvements when context is used, n=1-1 and 1-8). }
    \begin{subtable}{.5\linewidth}
        \centering
        %\caption{}
        \setlength{\tabcolsep}{0.6pt}
        \resizebox{0.95\textwidth}{!}
        {
        \renewcommand{\arraystretch}{1}
        \begin{tabular}{c|cc|cc|cc|cc}
        \hline
        %& \multicolumn{8}{c}{Binary Target}\\\cline{2-9}
        {\it n} & \multicolumn{2}{c|}{PromptOnly} & \multicolumn{2}{c|}{FinetunePrompt} & \multicolumn{2}{c|}{ClassHead} & \multicolumn{2}{c}{ClassLoRAHead}  \\\cline{2-9}
        & FRR $\downarrow$ & FAR $\downarrow$ & FRR $\downarrow$ & FAR $\downarrow$ & FRR $\downarrow$ & FAR $\downarrow$ & FRR $\downarrow$ & FAR $\downarrow$ \\\hline\hline
        1 & {\bf 13.8} & {\bf 46.7} & 12.6 & 16.4 & 20.6 & 2.0 & 19.5 & 2.4 \\
        8 & 37.2 & 56.7 & 18.6 & 2.3 &  17.7 & 1.6 & 17.5 & 2.2 \\\hline
        1-1 & 20.1 & 74.7 & 18.6 & 2.3 & 16.3 & 2.0 & 17.6 & 2.2 \\
        1-8 & 65.2 & 33.2 & {\bf 17.7} & {\bf 2.1} & {\bf 16.2} & {\bf 2.0} & {\bf 14.9} & {\bf 1.8} \\\hline
        \end{tabular}
        }
    \end{subtable}%
    \begin{subtable}{.5\linewidth}
        \centering
            %\caption{}
            \renewcommand{\arraystretch}{1}
            \setlength{\tabcolsep}{6pt}
            \resizebox{0.85\textwidth}{!}
            {
            \centering
            \begin{tabular}{c|ccc|ccc}
            \hline
             & \multicolumn{3}{c|}{ClassHead} & \multicolumn{3}{c}{ClassLoRAHead}  \\\cline{2-7}
            \textit{} & \text{EER} & \text{FAR} & \text{FAR} & \text{EER} & \text{FAR} & \text{FAR}\\
            \textit{n} & \text{\%} & \text{@5\%} & \text{@10\%} & \text{\%} & \text{@5\%} & \text{@10\%}\\
            \textit{} & \text{} & \text{FRR} & \text{FRR} & \text{} & \text{FRR} & \text{FRR}\\\hline\hline
            1   & 10.5& 29.9 & 11.4 & 10.6 & 30.1 & 12.7\\\hline
            8   & 9.5 & 25.6 & 8.3 & 9.6 & 30.5 & 8.7\\\hline
            1-1 & 8.5 & 18.3 & 6.1 & 8.9 & 19.9 & 7.3\\\hline
            1-8 & {\bf 7.9} & {\bf 16.7} & {\bf 4.8*} & {\bf 8.5} & {\bf19.5} & {\bf 6.1*}\\
            \hline
            \end{tabular}
            }
    \end{subtable}    
    \label{table:results}
\end{table}

\vspace{-4mm}
\subsection{LLM Finetuning}
\label{llm:finetune}
We use a pretrained instruction-tuned LLM, Vicuna-7B-v1.3~\cite{vicuna2023} in the experiments. The inference is done using 4 NVIDIA A100 GPUs. For finetuning the prompting-based approach, we train parameters of the LoRA adapters~\cite{hu2021lora} for 3 epochs using 8 GPUs with a learning rate of 2e-5, with a warmup to over 3\% of the learning steps. The LoRA adapters (rank=8) have 4.1M parameters that is only 0.06\% of the 7B parameters of the used LLM. In the classification-based approach, we experiment with the full finetuning, and with LoRA adapters. We experimented with different number of training epochs, and we trained the former for 1 epoch, while for LoRA we use 3 epochs, as those were the best on the validation set (we found it to be less prone to overfitting). The classifier head adds only $\sim$8k additional parameters (the rest are the same as in the prompting-based approach). We use FastChat toolkit~\cite{zheng2023judging} for inference and finetuning with DeepSpeed GPU optimization~\cite{rasley2020deepspeed}, with $0$ temperature during inference. Finetuning is done on \textit{train/val} parts of the used dataset.

\vspace{-2mm}
\subsection{Evaluation Procedure}
\label{llm:eval_proc} 
We evaluate the proposed approaches for the following setups. Firstly, we quantify the impact of the ASR uncertainty on classification of the follow-up query only (thus, no context), by representing it with $1$- and $n$-best hypotheses. We set $n=8$, found to work best on the validation data. Secondly, we quantify the impact of adding the context via the initial query, for which we only consider $1$-best hypothesis (for the follow-up we compare both). We compare both the prompting- and classification-based approaches (see Table~\ref{table:results}). 
Since the prompting-based approach produces only the textual binary outputs (``0'' or ``1'' tokens), we compare in terms of False Accept Rate (FAR) and False Reject Rate (FRR) metrics, where the lower these metrics the higher the system accuracy. However, this approach does not enable to tune the DDSD system in practical applications as we cannot set target operating points (OP) (the binary labels do not allow to define the regime under a desired FRR). On the other hand, the classification-based approach is tunable as it provides probability scores. We report Equal Error Rate (EER) and FAR at hypothetical OPs 5\% and 10\% FRR. This allows us to establish the model accuracy across regimes relevant for user experience. 

%\vspace{-4mm}
% \subsection{Baseline Systems}
% \label{sec:baselines}
% \vspace{-2mm}
% For the DDSD task, we use the LatticeRNN model~\cite{jeon2019latrnn} as our baseline. LatticeRNN is a strong baseline system as it processes the whole ASR decoding lattice whereas $n$-best list used in proposed ASR+LLM approach is a condensed prompting-friendly feature derived from the lattice. We also use the LatticeRNN to \textit{``teach''} the ASR+LLM system to output its decision on a scale of 0 to 100 and LatticeRNN could also be seen as a \textit{teacher} model under the \textit{student-teacher} knowledge-transfer paradigm. For the KS task, we devise a trivial baseline which claims that a keyword was detected if the ASR outputs that keyword as the 1-best hypothesis. If the ASR 1-best hypothesis does not match one of the 10 command keywords, we classify the utterance as OOV. We expect our ASR+LLM approach to improve upon this trivial baseline by rectifying the ASR predictions using the information in the $n$-best list prompts.

\vspace{-2mm}
\section{Results and Discussion}
\label{results}

The prompting-based approach exploits the generative power of the LLM to \textit{understand} the task based on the {\it prompt}, and outputs an answer that is then parsed to obtain the binary prediction for the speech being device-directed. We prompt the LLM system with the {\it task-prompt} and the {\it utterance prompt}, as described in Sec.\ref{prompt_base}, without any finetuning (``PromptOnly''). We confirm empirically that the LLM can indeed perform the task out-of-the-box, but sometimes it outputs a descriptive answer in natural language, as opposed to the binary output (as desired for the task). To deal with this, when the LLM outputs a descriptive answer different from ``0''/``1'' tokens, we consider the follow-up utterance to be a device-directed speech (instead of removing the target utterance, which affects $\sim$6\% of our data). The finetuned LLM (``FinetunePrompt'') outputs only ``0''/``1'' values.

Table~\ref{table:results} ({\it left}) compares the prompting- vs. classification-based approaches. When prompting the LLM without finetuning (``PromptOnly''), the model performs best with the simple $n=1$ hypothesis and without context. This is expected since the model is not finetuned on the target task, also suggesting that the LLM is not familiar with the concept of $n$-best lists from its original training, and, thus, it is unable to leverage the ASR uncertainty effectively (using the prompt design proposed here). The model becomes confused by additional information (coming either from the ASR uncertainty in the follow-up prompt, and/or context from the initial query), and such a system is not useful in practice (note the high FAR of 46.7\%). On the other hand, when we LoRA finetune the prompt-based model (``FinetunePrompt'') to output binary targets, the accuracy improves considerably. Note a reduction in FAR: when using the $1$-best hypothesis, FAR reduces 46.7\% $\rightarrow$ 16.4\%. Interestingly, by using the $n=8$-best hypotheses (without context) brings the FAR to as low as 2.3\%. While this comes at the expense of the increase in FRR (12.6\% $\rightarrow$ 18.6\%), it still renders a model that can be useful in practice. By looking at the role of the context, we note that adding the $n$-best hypothesis for the initial query helps when the $n$-best hypothesis is used for the follow-up (FAR 16.4\% $\rightarrow$ 2.3\%). This shows the importance of the context. However, the cumulative improvement in accuracy is reduced when both the context and ASR uncertainty are used jointly in this experiment. We also observed diminishing gains as we increase the size of the $n$-best list beyond $n=8$, which can be due to the model overfitting and/or getting more confused by the uncertainty from too many ASR hypotheses. As observed in~\cite{dighe2024llm}, LoRA finetuned prompting-based approach (``FinetunePrompt'') does not require a task prompt since the model focuses on the target label, given the utterance prompt. On the other hand, the full classification finetuning was robust to whether the prompt was used or not in this task.

Next, we compare the classification-based approaches (``ClassHead'' and ``ClassLoRAHead''), where a linear classifier (head) is added on top of the base LLM (see Sec.~\ref{class_base}). To compare with their prompting-based counterparts, for fairness, we binarize the classifier probability scores with 0.5 threshold. First, we see that ``ClassHead'' is able to leverage the ASR uncertainty (no context), with reduction in FAR (2.3\% $\rightarrow$ 1.6\%). By adding context, we note the reduction in FRR, however, it increases FAR (1.6\% $\rightarrow$ 2.0\%). Overall, these results suggest that the full finetuning of the classifier on top of the LLM does not benefit from additional context in this regime (with the used threshold). However, despite similar accuracy to that of ``FinetunePrompt'', the ``ClassHead'' approach is still a preferred candidate for the real-world usage as it can be tuned to achieve desired user experience (as it outputs probabilistic scores). Note that the LoRA adapters in the classification-based approach (``ClassLoRAHead'') are able to efficiently use both, the ASR uncertainty and previous context, with the best achieved FRR of 14.9\% and FAR of 1.8\%. This approach is also easier and faster to train, as only a fraction of the model parameters need to be updated. 

Table~\ref{table:results} ({\it right}) shows the results by the classification-based approach at different regimes using its probabilistic output. Specifically, we report for two exemplary OPs: FRR of 5\% and 10\%, and measure the EER/FAR that the system would produce under these false rejections, allowing us to directly compare the two classification models (``ClassHead'' and ``ClassLoRAHead''). Overall, ``ClassHead'' approach is more accurate, as it allows to tune all model parameters. The results support our hypothesis that modeling both, the ASR uncertainty and context of the previous query, {\it jointly} is critical for improving the model accuracy ($\sim$20-40\% reduction in FAR at 10\%FRR, when comparing the rows with $n$=$8$ and $n$=$1$-$8$). To better understand the models' behaviour, we listened to a number of utterance pairs that were falsely accepted at 10\% FRR by models without context modeling (rows 2-3), but rejected when context is used (rows 4-5). Since there was no a continuation of the first request, the context (i.e. the initial request) helped to disambiguate the decision for the follow-ups. We observed a similar trend for the false rejects, where the context helped to accept the follow-ups as device-directed speech. To quantify the overall robustness, we report Detection Error Trade-off (DET) curve in Supp. Materials.

\vspace{-4mm}
\section{Conclusions}
\label{sec:conclusions}
\vspace{-4mm}
We explored various techniques for adapting an LLM model, including prompt tuning, and fine-tuning strategies. We showed that by integrating knowledge from a previous request of the user can largely improve the accuracy of device-directed speech detection in the follow-ups. These are often unconstrained in the content, and may not always be directed to the device, therefore, classifying them accurately is important for enabling naturalistic user experience and engagement. Specifically, we showed that the prompting-based approach leads to high accuracy when the LLM is fine-tuned on the DDSD task, however, it exhibits poor performance when {\it only} the engineered prompts are used (despite of adapting the language content in the prompts). On the other hand, adding a classifier on top of the pretrained (frozen) LLM allows the model to focus on the contextual clues from the first (initial) query, as well as to leverage effectively the ASR uncertainty (via the n-best hypotheses) -- both encoded in the prompts, when making decisions about the follow-ups. We also showed that the classifier-adapted approaches can achieve high accuracy by tuning a fraction of the LLM parameters (via LoRA), saving the compute time. While in this work we focused on the pairs of the user queries as input to the LLM, other signals (e.g. VA's responses, acoustic features, and speaker information) can also be integrated to further improve the system accuracy. 
%This is part of our ongoing research. Finally, we showed that this approach is more practical for the systems where decisions need to be made within a desired operating region (that is application-dependent). 
\vspace{-4mm}
\footnotesize\bibliography{refs}
\bibliographystyle{IEEEtran}
\appendix
\pagebreak
\section{Supplemental material}
\begin{figure}[h]
    \centering
    \includegraphics[width=0.8\linewidth, height=0.6\linewidth]{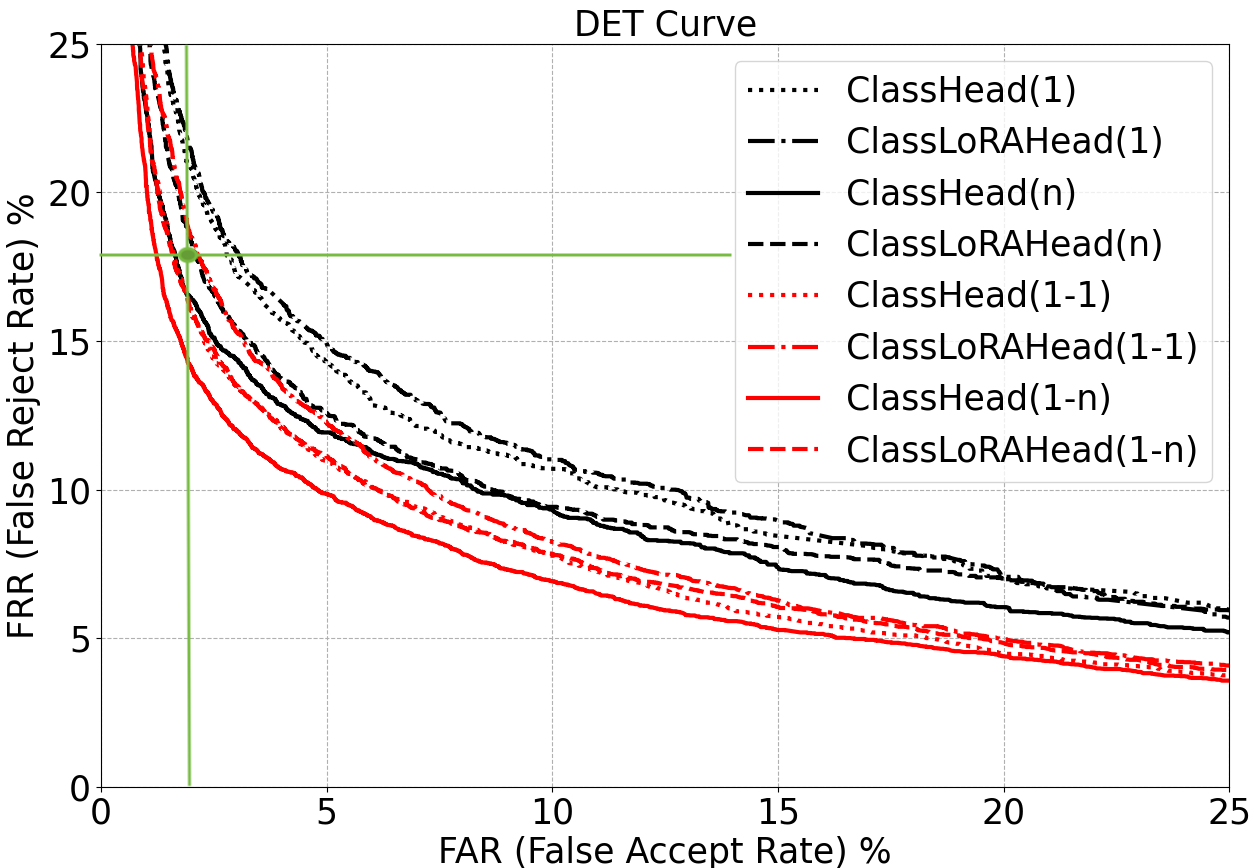}
    \caption{DETs showing the accuracy of the models across a range of regimes. Next to the model names in the brackets, we show the type of ASR hypothesis for each model reported ($1$- and $n=8$,w found on the validation set to work best for the task). The lines in {\it black} and {\it red} represent the models without and with context of the previous query, respectively. The {\it green} dot shows the accuracy of the best prompting-based approach (``FinetunePrompt'' with LoRA adapters).}
    \label{fig:roc_ddsd}
\end{figure}

To quantify the overall robustness of the models, we report Detection Error Trade-off (DET) curve in Fig.~\ref{fig:roc_ddsd}. We see a similar trend across most of the regimes, with both ASR uncertainty and context of the previous query in the classification-based approach performing the best, and also outperforming the prompting-based approach (``FinetunePrompt'' with LoRA adapters, depicted in {\it green}). Finally, while the LoRA-based approach regresses compared to the full finetuning, with the largest regression occurring at 10\% FRR, where FAR goes from 4.8\% to 6.1\%, both models still achieve relatively high accuracy. The LoRA-based approach has less tunable parameters and is much faster to train (on the used dataset, the reduce in the train time was $\sim$5-10 times). Therefore, a trade-off needs be made depending on the hardware and time resources.

\end{document}